\documentclass[prl,twocolumn,superscriptaddress,showpacs,floatfix,amsfonts]{revtex4}
\usepackage{graphicx,graphics,color,epsfig}
\usepackage{bm}
\usepackage{amsmath}
\usepackage{amssymb}
\usepackage{epstopdf}
\begin{document}
\preprint{}
\title{Possible Kondo effect in the iron arsenides}
\author{Jianhui Dai}
\author{Guanghan Cao}
\affiliation{Department of Physics,  Zhejiang University, Hangzhou
310027, China}
\author{Hai-Hu Wen}
\affiliation{National Laboratory for Superconductivity, Institute of
Physics, Chinese Academy of Science, Beijing 100190, China}
\author{Zhuan Xu}
\affiliation{Department of Physics,  Zhejiang University, Hangzhou
310027, China}

\begin{abstract}

The normal state of the iron arsenides shows the poor metallic
behavior mixed with strong magnetic fluctuations. In particular,
some FeAs-1111 and FeAs-122 compounds show the linear-$T$ dependence
of susceptibility above the spin-density wave (SDW) transition and
the logarithmic upturn of resistivity at low temperatures. We
suggest that this is due to the spin-flip scattering between the
charge carriers and the local moments in the undoped FeAs layer
where Kondo effect coexists with the SDW. This scenario is also
accounted for the change of the magnetoresistance from positive to
negative in the Sr$_3$Sc$_2$O$_5$Fe$_2$As$_2$ compound.
\end{abstract}
 \maketitle

\narrowtext

{\it Introduction.} The recent discovery of superconductivity in the
iron-arsenide materials with transition temperatures up to 56~K
poses a new intellectual challenge in the correlated electron
systems
\cite{Kamihara:JACS08,XHChen:08,GFChen:08,Ren:08,Wen:08,Wang:08}.
These materials consist of an unique two-dimensional conducting FeAs
layer. The parent quaternary compounds (the 1111 series), such as
LaFeAsO, show a structure distortion and a collinear spin-density
wave (SDW) transition. They are suppressed by electron doping,
giving rise to the superconductivity\cite{Cruz:08}. This is
remarkably similar to the cuprates, where the high temperature
superconductivity is derived from the magnetic ordered parent
compounds by doping electrons or holes into the CuO$_2$ planes.

An apparent difference between the iron arsenides and the cuprates
is that the parent compound of the former is a multi-orbital metal.
Yet the normal state properties of the FeAs-based materials are
unusual. One of the most striking feature is the linear temperature
dependence of static susceptibility above the SDW transition
temperature. This behavior has been recently suggested as due to the
strong exchange interaction among the so-called {\it "preformed SDW
moments"} \cite{ZhangGM:08} or to the peculiarity of a two
dimensional Fermi liquid\cite{Korshunov:08}. Another silent feature,
perhaps of equal significance, is the resistivity upturn at low
temperatures in some 1111 series
with\cite{Kamihara:JACS08,GFChen:08,XHChen:08} or without
oxygens\cite{Wen2:08}. This feature is hindered by the emergent SDW
instability or superconductivity and is hidden in some ternary
compounds (the 122 series)\cite{Rotter:08,Sasmal:08}.

In our point of view, the above two features are characteristics of
the normal state of a single FeAs layer. The experimental evidence
for the resistivity upturn being robust and intrinsic to the FeAs
layer is accumulating. The transport measurement on a sizable single
crystal of the BaFe$_2$As$_2$ sample exhibits this behavior both in
the zero and finite magnetic field\cite{XHChen2:08}. Recently, a
noticeable resistivity upturn behavior was found to pertain above
$T_c$ in the superconducting LaFe$_{1-x}$Ni$_{x}$AsO
sample\cite{Cao:08}. Furthermore, a new 32522 compound, i.e.,
Sr$_3$Sc$_2$O$_5$Fe$_2$As$_2$, was found to show a significant
crossover from the metallic behavior at high temperatures to the
upturn feature at low temperatures\cite{Zhu:08}. The inter-layer
spacing in this compound is much larger than those in the 1111 and
122 series. So it serves as an ideal system to study the physics of
a single FeAs layer.

In this paper, we suggest that the two characteristic features can
be understood as due to the coexistence of Kondo effect and the SDW
in the undoped FeAs layer. This Kondo effect may be regarded as a
manifestation of the 3$d$ multi-orbital correlations and it can be
suppressed by electron doping and enhanced by small applied field.
We note that the resistivity upturn at low temperatures is usually
attributed to the charge localization or the Jahn-Teller effect. The
purpose of the present paper is to provide qualitatively evidence
that the Kondo scenario is more likely. Our interpretation of the
resistivity upturn assumes the existence of {\it "local moments"}
\cite{Dai:08}, while the latter is evidenced in LaFeAsO by recent
electron spin resonance experiment\cite{Wu:08}.

{\it Susceptibility and resistivity.} Let us first briefly discuss
some experimental results. The linear temperature dependence of the
susceptibility above the SDW ordering temperature is universal in
the iron arsenides including the F-doped 1111 \cite{Klingeler:08}
and 122 series\cite{XHChen2:08,GWu:08}. This behavior is an
indication of strong antiferromagnetic interaction among the local
moments, described by the $J_1$-$J_2$ Heisenberg model\cite{Si:08,
Ma2:08}. If one chooses $S=1$ and  assumes $J_1=J_2 (\equiv J_d)$,
the susceptibility can be fitted by $\chi=\chi_0 (1+a~T)$ (for
$T<0.3 J_d$)\cite{ZhangGM:08}. For LaFeAsO$_{1-x}$F$_{x}$ or
LaFe$_{1-x}$Ni$_{x}$AsO, however, while the coefficient of the
linear-$T$ term is nearly independent on $x$, the Pauli part,
$\chi_0$, decreases with increasing $x$~\cite{Cao:08}.

For the resistivity, a noticeable upturn just before the SDW
transition could be seen in LaFeAsO [the upper inset in Fig.~1(a)].
This behavior can be identified even in the normal state of the
optimally-doped superconducting samples of
LaFe$_{1-x}$Ni$_{x}$AsO~\cite{Cao:08} [the upper inset in
Fig.~1(b)]. The upturn after the minimum can be fitted by $R\propto
c_1-c_2\log T$ [Fig.~1(a,b)]. Moreover, the resistivity crossovers
to $ R(T)\propto R(0)[1-cT^2]~$ as $T$ decreases further [the lower
inset in Fig.~1(b)]. Such kind of resistivity evolution is a
benchmark of Kondo effect\cite{Hewson}.
\begin{figure}[ht]
\epsfxsize=8.0cm \centerline{\epsffile{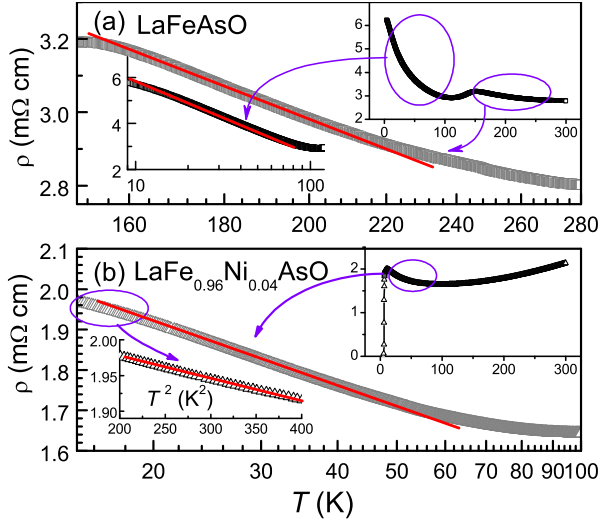}}
\caption[]{(Color online) Temperature dependence of the resistivity
in the LaFeAsO and LaFe$_{0.96}$Ni$_{0.4}$AsO samples. The
temperature axis is scaled in $\log T$, while it is scaled in $T^2$
in the lower inset in Fig.~1(b). The straight red lines are guide to
the eyes. The experiment data are taken from Ref.\cite{Cao:08}.}
\label{fig:32522}
\end{figure}

{\it The spin-flip scattering and the local Fermi liquid.} We now
argue that the linear-$T$ susceptibility and the Kondo-like
resistivity can be understood as due to the co-existence of the
local moments (${\vec S}_{j\alpha}$) and itinerant carriers
($c_{{\bf k}\alpha\sigma}$) in the FeAs layer\cite{Dai:08}. In the
LDA language\cite{Singh:08}, they correspond to the coherent
excitations in the vicinity of the Fermi-level and the incoherent
excitations away from the Fermi-level, as illustrated in Fig.~2(a).
The incoherent part is modeled by the $J_1$-$J_2$ Heisenberg model:
$ H_{J}=J_1\sum_{(n.n.)\alpha}{\vec S}_{i,\alpha}\cdot{\vec
S}_{j,\alpha}+J_2\sum_{(n.n.n.)\alpha}{\vec S}_{i,\alpha}\cdot{\vec
S}_{j,\alpha}+J_H\sum_{j,\alpha,\beta}{\vec S}_{j,\alpha}\cdot{\vec
S}_{j,\beta}$. The coherent part is modeled by non-interacting
fermions with hole and electron pockets near the $\Gamma$ and $M$
points respectively. The interaction between them is generally
described by
\begin{eqnarray}
\sum_{{\bf k}{\bf
q}\alpha\beta\sigma\sigma'\gamma}\sqrt{w_{\alpha}w_{\beta}}G_{{\bf
k,q}\alpha\beta\gamma} c^{\dagger}_{{\bf
k+q}\alpha\sigma}\frac{{\vec \tau}_{\sigma\sigma'}}{2}c_{{\bf
k}\beta\sigma'}\cdot {\vec S}_{{\bf q}\gamma},\nonumber
\end{eqnarray}
where $w_{\alpha}$ is the weight of the coherent excitations in the
$\alpha$-th orbital\cite{note1}. So the system may be regarded as a
generic momentum resolved Kondo lattice. Some components of $G_{{\bf
k,q}\alpha\beta\gamma}$ may be {\it antiferromagnetic} due to the
strong orbital-mixing and the moderate 3$d$-Coulomb interaction,
surpassing the Hund's coupling. Though all five 3$d$-orbitals
contribute to $w$, some of $w_{\alpha}$ may be vanishingly small due
to possible orbital-selective Mott transition [Note that
$w=\sum_{\alpha}w_{\alpha}$ is small for a bad metal]. With these in
mind, we consider a simplified model with a single
conduction-electron band:
\begin{eqnarray}
H=w\sum_{{\bf k}\sigma} (E_{{\bf k}}-\mu) c^{\dagger}_{{\bf
k}\sigma}c_{{\bf k}\sigma}+w J_K \sum_{j\alpha}{\vec s}_{j,c}\cdot
{\vec S}_{j,\alpha}+H_J.
\end{eqnarray}
Where, $E_{{\bf k}}$ is the band energy with the width $2D$ [as
illustrated in Fig.~2(a)], $\mu$ is the chemical potential, ${\vec
s}_{j,c}$ is the spin operator of the charge carrier, $J_K$ the
Kondo coupling and $\mu$ the chemical potential.
\begin{figure}[ht]
\epsfxsize=7.0cm \centerline{\epsffile{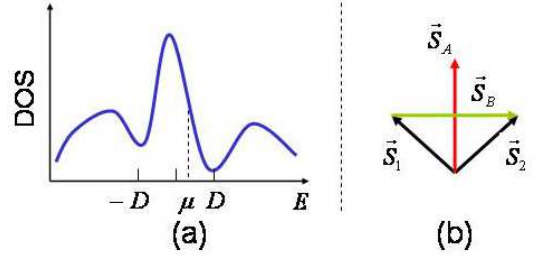}}
\caption[]{(Color online) (a) Schematic picture for coherent and
incoherent excitations. (b) The hybridization (red) and
non-hybridization (green) components distinguished in the MF
approximation.} \label{fig:illustrate}
\end{figure}

For the simplest case with $\alpha=1$, i.e., the $S=1/2$ Kondo
lattice model, the coexistence of Kondo effect and antiferomagnetic
ordering is possible for suitable $J_K/D$ and
$J_d/(wD)$\cite{Si:06,Vojta:08}. A more plausible case is for
$\alpha=1, 2$. This case is similar to the underscreened $S=1$ Kondo
lattice with an additional Hund's coupling. It can be solved by the
mean field (MF) approximation \cite{Perkins:07}, and the coexistence
of Kondo effect and magnetic ordering arises naturally in the MF
solution. The ground state can be intuitively understood by
distinguishing the hybridization and non-hybridization components
[see in Fig.~2(b)], defined by ${\vec S}_{A}={\vec S}_{1}+{\vec
S}_{2}$ and ${\vec S}_{B}={\vec S}_{2}-{\vec S}_{1}$ respectively.
The distinction between the two components becomes more
apparent if there is a weak Ising anisotropy. So the local moments
are partially screened by the charge carriers, leading to the
residual effective spins at each site. The magnitude of each of them
decreases with increasing $J_K$ and arrives at the minimal value
$1/2$ if the hybridization sector is in the Kondo phase. The local
magnetic entropy is suppressed further by the $J_1$-$J_2$
frustrations, leading to a collinear SDW with a reduced ordered
moment $m_d$ observed experimentally.

In the Kondo phase of both cases, the low energy physics is
determined by the Kondo temperature, $T_K$, and the coherent
temperature, $T_{coh}$. The ratio of them can be approximated by
$T_{coh}/T_K\sim e^{-f(\mu)}$,
$f(\mu)=\int^{D-\mu}_{-D-\mu}[\frac{1}{2}-\frac{\rho(\mu+\omega)}
{\rho(\mu)}]\frac{d\omega}{\omega}$\cite{Burdin:00}. $T_{coh}/T_K$
decreases exponentially if there is a drastic variation in
$\rho(\omega)$ near the chemical potential $\mu$, particularly when
$\mu$ moves to the band edge~[see in Fig.~2(a)]. The LDA+DMFT
calculation shows that $T_{coh}$ is also strongly suppressed by the
Hund's coupling~\cite{Haule:08}. For the same reasons, the specific
heat coefficient is suppressed. Thus for the iron arsenides we shall
focus on the limit $T_K\gg T_{coh}\approx 0$. In this limit, the low
temperature transport property is governed by a single scale $T_K$.
Therefore, the normal state is a local Fermi liquid. Here the bad
metal behavior is also featured by comparably small ratio of the
saturated resistivity to the one at room temperature (say at $T=300$
K) due to small $w$. In the high temperature regime, however, the
transport may deviate from the Fermi liquid behavior due to
fluctuations beyond the MF approximation or other neglected
multi-orbital effects.

{\it Implications of Kondo effect for the 1111 and 122 series.} The
local Fermi liquid solution provides a scenario for the coexistence
of Kondo effect and collinear SDW. The true long range ordering can
be established at finite temperatures with small inter-layer
coupling $J_z$. The ordering temperature $T_{SDW}\approx 4\pi
\Delta_{SDW}/\ln(\Delta_{SDW}/J_z)$, with $\Delta_{SDW}=J_d
m_d$\cite{Xu:08}.

In the absence of $J_K$, the static susceptibility is contributed
from both types of excitations: $\chi=\chi_{inc}+\chi_{coh}$. Above
$T_{SDW}$, $\chi_{inc}$ is given by the $J_1$-$J_2$ model,
$\chi_{inc}={\tilde \chi_0}(1+aT)$ for $T<0.3J_d$\cite{ZhangGM:08}.
$\chi_{coh}$ is contributed from the coherent excitations, which is
of the Pauli form. This part is proportional to the DOS at the Fermi
level, $\rho_F$. So we have $ \chi=\chi_0+\alpha T, $ where
$\chi_0=\chi_{coh}+{\tilde \chi}_0$, $\alpha=a{\tilde \chi}_0$.
According to the LDA calculation, the electron doping (either via F
or Ni) will shift $\mu$ to the right, leading to a decrease in
$\rho_F$, as illustrated in Fig.~2(b). Thus the Pauli part $\chi_0$
shows an overall decreasing tendency, in agreement with the
Ni-doping experiment\cite{Cao:08}. With the finite $J_K$, as $w$ is
very small for a bad metal, the effective coupling $wJ_K$ does not
essentially change the linear behavior above $T_{SDW}$ but may
slightly enhance the Pauli susceptibility below $T_K$. As a peculiar
feature of the coexistence of the Kondo effect and the $J_1$-$J_2$
coupling, the local moments do not show a significant Curie-like
behavior for $T\gtrsim T_K$ as in the conventional Kondo effect.

The low temperature upturn behavior of electronic resistivity is
usually attributed to the localization effect driven by
doping-induced disorders. But this is less likely in the iron
arsenides, as the upturn appears most significantly in the parent
compounds and decreases with increasing Ni-doping. We emphasize that
the electron correlation could be responsible to the upturn behavior
in a translational invariant system. In the underdoped cuprates, the
strong Coulomb repulsion results in a gauge interaction between the
spinon and the holon, leading to a metal-insulator transition in the
pseudo-gap phase\cite{Marchetti:07}. In the iron arsenides, the
Coulomb repulsion is moderate so that the normal state is a bad
metal which was suggested to be in proximity to the Mott
transition\cite{Si:08}. Here we just interpret the incoherent
transport as due to the coexistence of Kondo effect and the SDW.
Because $T_{coh}\approx 0$, the normal state is characterized by the
Kondo temperature $T_K\sim D e^{-1/(wJ_K \rho_F)}$. For $T\lesssim
T_K$, the resistivity follows the well-known logarithmic behavior:
$R\approx c_1-c_2\log T$, with the coefficient $c_2\sim wJ_K
\rho_F$. This explains why $c_2$ and $\chi_0$ show the same overall
doping dependence in the LaFe$_{1-x}$Ni$_x$AsO
compound\cite{Cao:08}.

When $T$ goes down to $T_{SDW}$, the SDW ordering gap $\Delta_{SDW}$
opens, triggering the SDW instability of the charge carriers. So
$\rho_F$ has a drop across $T_{SDW}$. As a result, $\chi_0$ has a
drop too as observed in some 122 series\cite{XHChen2:08}.
Accordingly, there are two different Kondo temperatures, $T^{(1)}_K$
and $T^{(2)}_K$, associated with $\rho_F$ for $T>T_{SDW}$ and
$T<T_{SDW}$, respectively. The resistivity has also a drop due to a
drastic increase in the scattering rate $1/\tau$ at the nested Fermi
surfaces. So if $T^{(1)}_K\sim T_{SDW}$, there is a resistivity
upturn above the SDW transition (as in many 1111 series). Otherwise,
for $T^{(1)}_K\ll T_{SDW}$, no upturn appears above $T_{SDW}$ (as in
some 122 series). These two cases are in accordance with the fact
that $J_z$ decreases from the 122 to 1111 series. In the SDW phase,
the upturn shows up again when $T\lesssim T^{(2)}_K$[as in
Fig.~1(a)]. The feature is most significant if $T_K\gtrsim T_{SDW}$
or $\Delta_{SDW}\rightarrow 0$. This is the case for the 32522
sample (see the inset in Fig.~3) where $J_z$ should be very small.
\begin{figure}[ht]
\epsfxsize=7.0cm \centerline{\epsffile{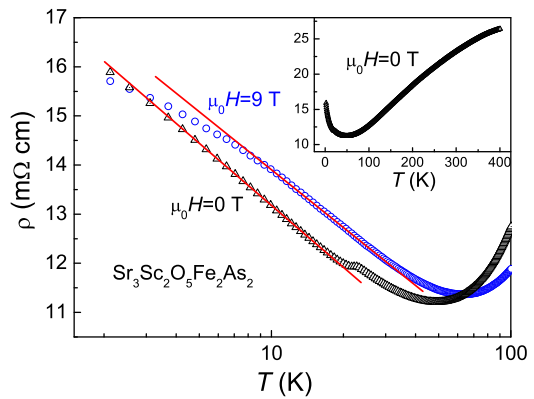}}
\caption[]{(Color online) Temperature dependence of the resistivity
for the 32522 sample. The temperature axis is scaled in $\log T$.
The straight red lines are guide to the eyes. The experiment data
are taken from Ref.\cite{Zhu:08}.} \label{fig:32522}
\end{figure}

{\it The magnetoresistance and the possible ordered moment in the
32522 sample.} At zero temperature the magnetic field dependence of
the resistivity, the magnetoresistance (MR), can be deduced from the
Friedal sum rule\cite{Hewson}. In the conventional Kondo problem,
the impurity spin is completely screened so that the MR is negative.
This is because the spin-flip scattering and spin compensated ground
state are suppressed in the presence of magnetic field.

However, the situation for the present Kondo effect is different.
Now the local moments are only partially compensated. The residual
ordered moment plays the role of an intrinsic magnetic field. It is
in general a function of the external magnetic field $h$, denoted by
$m_d^{(h)}$. Note that $m_d^{(h)}$ will be always suppressed by $h$
for $h<h_{1}$, with $h_{1}$ being defined by $m_d^{(h_{1})}=0$. For
$h>h_{1}$, the local moment polarizes gradually along $h$ and
saturates at the original quantized magnitude ( e.g., 0.5 in the
unit of $g\mu_B$ ) as $h$ approaches $J_d$. Applying the sum rule
gives contribution of the local moment to the
resistivity:~$R_{loc}(m_d^{(h)})=R_{loc}(0)\cos^2(\pi m_{d}^{(h)})$,
with $R_{loc}(0)$ being the zero temperature resistivity for $h=0$
or $m_d^{(0)}\equiv m_d$. So we have
\begin{eqnarray}
\frac{\Delta R_{loc}(h)}{R_{loc}(m_d)}=\frac{\sin\pi (m_d-m_d^{(h)})
\sin\pi(m_d+m_d^{(h)})}{\cos^2(\pi m_d)}.
\end{eqnarray}
When $h>h_{2}$, where $m_d^{(h_{2})}=m_d$, the sign of the MR
changes from positive to negative.

We find that the MR result depends only weakly on the details of
$m_d^{(h)}$ if $m_d$, $h_{1}$, and $J_d$ are properly fixed. Thus we
can use a simplified molecular field approximation, i.e.,
$m_d^{(h)}= m_d(1-h/h_1)$ for $h<h_1$ and
$m_d^{(h)}=0.5\sqrt{\frac{h-h_1}{J_d-h_1}}$ for $h_1<h \leq J_d$.
Eq.(2) is then compared with the $T=2$~K data in the 32522 sample.
Fig.~4 shows the MR result of this estimation with $m_d\approx 0.014
~g\mu_B$, $h_1\approx 1.15$ Tesla, and $J_d\approx 60$ meV. We
expect that the estimated MR curve could be more smooth if
$m_d^{(h)}$ is accurately calculated from the $J_1$-$J_2$ or O(3)
models in the presence of external field $h$. It is interesting to
note that the very small ordered moment obtained by fitting the MR
data may be responsible to a tiny kink at zero field seen in Fig.~3
near $T\approx 20$~K. The kink disappears at $h=9$ Tesla when the
SDW gap ($\approx 0.8$ meV) is completely suppressed. Our
observation also suggests that both the positive and negative parts
of the MR should be robust as $T\rightarrow 0$.
\begin{figure}[ht]
\epsfxsize=7.0cm \centerline{\epsffile{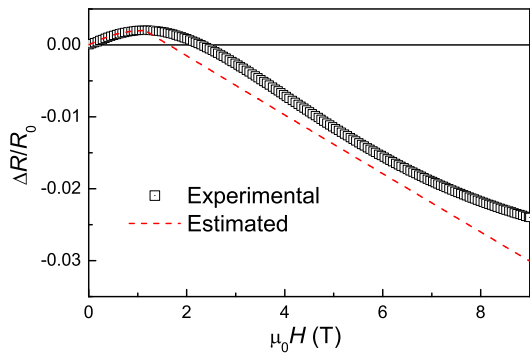}}
\caption[]{(Color online) The  magnetoresistance at $T=2$~K for the
32522 sample. The experimental data (open squares) are taken from
Ref.\cite{Zhu:08}. The red dashed line is a theoretical estimation
at zero temperature based on the Friedal sum rule.} \label{fig:MR}
\end{figure}

Finally, we remark that for the undoped 1111 and 122 compounds,
$\Delta_{SDW}\approx 15-25$ meV, so $h_1\sim 150-250$ Tesla. This
means that the negative MR is hardly observable experimentally. Upon
electron doping or chemical pressure, however, $\Delta_{SDW}$
decreases dramatically, the negative MR could be observed if the
superconductivity does not intervene the Kondo effect. The chemical
pressure can be realized by P doping of As which effectively
enhances $w$ and suppresses $m_d$\cite{Dai:08} so that $T_{coh}$ may
increase slightly. All these could be possibly seen in the iron
phosphides where the superconductivity is weak and can be completely
suppressed by small magnetic field.

In summary, the logarithmic-$T$ resistivity behavior, its relation
with the doping dependence of the Pauli susceptibility, and the sign
change in the MR, indicate a possible Kondo effect in the single
FeAs layer. The coexist of Kondo effect and the SDW is inherited
from the multi-orbital 3$d$-electron correlations and plays a role
in the rich complexities of the iron arsenides.

We thank W.Q. Chen, X.H. Chen, Y. Liu, Z.Y. Lu, Q. Si, N.L. Wang,
Z.Y. Weng, H.Q. Yuan, F.C. Zhang, and J.X. Zhu for useful
discussions. This work was supported by the NSF of China and PCSIRT
(IRT-0754) of Education Ministry of China.

\end{document}